\documentclass[11pt]{article}
\usepackage{moriond,epsfig}

\def\be{\begin{equation}}
\def\ee{\end{equation}}
\def\bea{\begin{eqnarray}}
\def\eea{\end{eqnarray}}

\begin{document}
\vspace*{4cm}
\title{
FERMION MASS IN $E_6$ GUT WITH DISCRETE FAMILY PERMUTATION SYMMETRY $S_3$}

\author{ S. Morisi }

\address{Dipartimento di Chimica Fisica ed Elettrochimica di Milano and INFN sezione di Milano}

\maketitle\abstracts{
A discrete symmetry S3 easily explains all neutrino data. However it is
not obvious the embedding of S3 in GUT where  all fermions  live in the
same representation. We show that embedding S3 in E6 it is possible to
make distinction between neutrinos and the rest of matter fermions.
}

\section{Introduction}
So far it is not clear how to extend the standard model to include fermion masses.
In general the mass terms are arbitrary $N_f\times N_f$ complex matrices
$M_f$ where $N_f$ is the number of generations. $M_f$ are not univocally
fixed by experimental data. To reduce the remaining arbitrariness flavour and
gauge symmetry are used in the model building. 
Gauge coupling unification, anomaly cancellation and
charge quantization are hints to consider Grand Unification (GU) models.
We assume a gauge symmetry $G_g$ acting vertically within
each generation and a flavour symmetry $G_f$ acting horizontally between 
different generations and we study the group $G_g\times G_f$. 

We can get information about the flavour symmetry $G_f$ from the observed 
mass and mixing fermions hierarchies.
First we consider the {\it lepton sector}. The three neutrino analysis 
\footnote{If MINIBOONE will not confirm LSND results,
neutrino data are compatible with only two mass difference and the number of
neutrinos is three.} \cite{Maltoni} 
is well compatible with the following mixing matrix \cite{Harrison:2003aw} 
\begin{equation}\label{eq1}
U_{HPS}
=\left(
\begin{tabular}{ccc}
$\sqrt{\frac{2}{3}}$&$\frac{1}{\sqrt{3}}$&$0$\\ 
$-\frac{1}{\sqrt{6}}$&$\frac{1}{\sqrt{3}}$&$-\frac{1}{\sqrt{2}}$\\ 
$-\frac{1}{\sqrt{6}}$&$\frac{1}{\sqrt{3}}$&$\frac{1}{\sqrt{2}}$
\end{tabular}\right)
\end{equation}
called tri-bimaximal, where the heaviest third neutrino is maximally
mixed between $\mu$ and $\tau$ flavours and $\theta_{13}=0$ while the second 
neutrino is equally mixed between $e,~\mu$ and $\tau$.
Recently discrete symmetry are studied to explain neutrino mixing (\ref{eq1}),
in particular $S_3 $ \cite{Harrison:2003aw} \cite{Grimus} \cite{Ma:2004xa} \cite{Caravaglios:2005gf} \cite{Caravaglios:2005gw}, 
$A(4)$ \cite{Ma:2004xa} \cite{A4} and $S_4$\cite{S4}.
In {\it quark sector} the three mixing angles are small, the only 
relevant angle is the 1-2 Cabibbo angle which is smaller than the 
1-2 and 2-3 leptonic angles and the mass hierarchy is strong.
Since quarks mixing is very small we have more information
and constrains on flavour symmetry $G_f$ from leptons, where the mixing is 
larger than quarks.
Quarks and leptons mixing and mass hierarchies are very different. The neutrino
masses can be degenerate and the mixing is large (\ref{eq1}), while
quark and charged lepton masses are strong hierarchy and the CKM mixing 
matrix is very close to the identity. Very different
hierarchy in quark and lepton sectors, could be a problem in GU models where
in general quark and lepton Yukawa couplings are related.
We consider the problem to reconcile different mass and mixing hierarchy
with lepton-quark symmetry in unified models. In literature there are at least
two class of solutions. 
One possibility is to extend the observed lepton symmetry 
to all fermions \cite{Matsuda:2005pq}.
Another possibility is that scalars couple differently to charged
fermion Yukawa from neutrino Yukawa, but there are very interesting
different possibility like the {\it screening} mechanism \cite{Lindner:2005pk}.
We have studied \cite{Caravaglios:2005gf} the possibility to make a 
difference between neutrino
and charged fermions selecting the gauge group $G_g$ and its scalar sector.

\section{Leptonic flavour symmetry}
In this section we study the flavour symmetry $G_f$ that follows from 
lepton sector explaining very well neutrino data,
then we will select the gauge group in the next section. 
Neutrino mass matrices $M_\nu$ is $\mu \leftrightarrow \tau$ invariant 
($S_2$ symmetric) only if it commutes with the matrix $P$
$$
\begin{small}
P=\left( 
\begin{array}{ccc}
1 & 0 & 0 \\ 
0 & 0 & 1 \\ 
0 & 1 & 0
\end{array}
\right),~~ P^{-1}~M_\nu~P=M_\nu
~~\Rightarrow~~
M_\nu=
\left(\begin{array}{ccc} 
a&d&d\\
d&b&c\\
d&c&b
\end{array}\right).  
\end{small}$$
Neutrino mass matrix $M_\nu$ and $P$ are diagonalized by the same unitary matrix $O$ which is
$$
\begin{small}
O(\theta)=\left( 
\begin{array}{ccc}
-\cos \theta & \sin \theta & {0} \\ 
\frac{1}{\sqrt{2}}\sin \theta & \frac{1\ }{\sqrt{2}}\cos \theta & {-\frac{1}{%
\sqrt{2}}} \\ 
\frac{1}{\sqrt{2}}\sin \theta & \frac{1}{\sqrt{2}}\cos \theta & {\frac{1}{%
\sqrt{2}}}
\end{array}
\right)\hspace{1cm}\Leftrightarrow \hspace{1cm}
\left(
\begin{tabular}{ccc}
$\sqrt{2/3}$&$\frac{1}{\sqrt{3}}$&{$0$}\\ 
$-\frac{1}{\sqrt{6}}$&$\frac{1}{\sqrt{3}}$&{$-\frac{1}{\sqrt{2}}$}\\ 
$-\frac{1}{\sqrt{6}}$&$\frac{1}{\sqrt{3}}$&{$\frac{1}{\sqrt{2}}$}
\end{tabular}
\right).
\end{small}
$$
Assuming the charged leptons mass matrix diagonal, $O$ is the PMNS leptonic
mixing matrix where the angle $\theta$ is the solar angle and it is not fixed by
the $\mu \leftrightarrow \tau$ symmetry while the atmospheric and $\theta_{13}$
angles are the same of the tri-bimaximal (\ref{eq1}). To obtain the solar angle 
we need that the singlet eigenstate $(1,1,1)$ is and eigenvector of the 
mass matrix $M_\nu$ and there are two possible solutions: {\it i)}
$M_\nu$ is $S_3$ invariant ($S_3$ is the permutation group of three objects)
or {\it ii)} the parameters in $M_\nu$ are constrained by $a=b+c-d$ which 
can follow directly from $A_4$ symmetry \cite{A4}. 
We are interested in the first case.
The $S_2$ permutation group is contained into $S_3$ and in general 
$S_3$ breaks spontaneously into $S_2$.  We have shown 
\cite{Caravaglios:2005gw} that in case $S_3\supset S_2$ breaking is {\it soft},
the solar angle is $\sin^2 {\theta}_{sol}=1/3 $ that agree very well with the 
experimental value. 
\section{A grand unification model for fermion masses}
In previous section we have said that in case $S_3$ symmetry is softly broken 
into $S_2$ ($\mu \leftrightarrow \tau$), we can explain very well neutrino mass 
and mixing hierarchies. Differently $S_2$ symmetry is strongly
broken in charged fermion sector. The issue is how to embed a neutrino mass 
matrix in the same unified gauge group where leptons and quarks Yukawa
are equal (lepton-quark unification). In SU(5) neutrinos and charged 
leptons Yukawa couplings are distinct since SU(5) does not contain
right-handed neutrino and we must introduce an additional singlet $S$  
\begin{equation}
L=g_{u}\,T^{\alpha \beta }\,T^{\gamma \delta }\,H^{\sigma }\,\varepsilon
_{\alpha \beta \gamma \delta \sigma }+g_{d}\,T^{\alpha \beta }%
F_{\alpha }\,\bar{H}_{\beta }+g_{v}\,\ F_{\alpha }\,S\,H^{\alpha
}+M S~S
\end{equation}
where $T$ and $F$ are the weyl fermions belonging to the 10 and $\bar{5}$ 
representation of SU(5).
However we are interested in model beyond SU(5) since
we want to explain why Yukawa are proportional to different vevs embedding 
SU(5) in bigger groups and we are interested in 
non supersymmetric extension of the Standard Model, but in such case gauge 
couplings do not unify in SU(5). Besides these general motivations, there
is one strong  reason to consider other gauge groups than SU(5).
Even if Yukawa couplings are distinct in SU(5), if we embed $S_3$ in SU(5)
requiring only one Higgs doublet
\footnote{We prefer to keep just one Higgs doublet , that will give
 mass both
for the up and the down sector. This is because we want to have the
 Standard
Model with just one higgs at \ the weak scale where the FCNC\ are strongly
suppressed due to the GIM\ mechanism.}, we obtain wrong prediction. In fact 
taking the Higgs as a $S_3$ singlet 
the only $S_3$ invariant renormalizable Yukawa operators in SU(5)
that give up mass terms are
$$
\lambda_1~ T_i~T_i~H~,~~~\lambda_2~T_i~T_j~H\label{tth}
$$
from which we obtain respectively the following up quark mass matrices
\begin{small}
\begin{equation}\label{mat1}
\lambda_1 ~v~\left(\begin{array}{ccc} 
1&0&0\\
0&1&0\\
0&0&1
\end{array}\right),~~~
\lambda_2~ v ~\left(\begin{array}{ccc} 
1&1&1\\
1&1&1\\
1&1&1
\end{array}\right)
\end{equation}
\end{small}
where $\lambda_1$ and $\lambda_2$ are arbitrary couplings and $v$ is the Higgs 
vev. The matrices (\ref{mat1}) give wrong masses ($m_u=m_c=m_t$) and mixings. The $S_3$ permutation 
symmetry is approximatively exact only for neutrinos
\footnote{We have proposed a 
phenomenological model \cite{Caravaglios:2005gw} where the Dirac neutrino 
mass matrix is proportional to the identity and mixing and mass hierarchies
come from Majorana mass terms that break softly $S_3$ into $S_2$ in the 2-3
direction through a seesaw mechanism.}, while the permutation symmetry is strongly broken in charged fermion sector.
Thus assuming only one Higgs doublet and the permutation symmetry,
the unifying gauge group choice is constrained by the fact
that the tree level Yukawa interactions are zero for charged fermions,
but this is not true for the SU(5) unification gauge group.
In the following we study one possible choice for the unifying group
so that only Dirac neutrino gets Yukawa coupling at tree level and
neutrino sector is distinct from charged fermion sector.  
If we embed SU(5) into SO(10) we have an additional U$_r$(1) gauge 
group that commutes with the full SU(5). 
The U$_r$(1) charges for the representation above are $H(+q),\bar{H}(-q),$
$T(-1)$ $F(+3)$ and $\nu_{R}(-5)$ \cite{slansky}. From these charges we derive that \ each
mass operators have U$_{r}$(1) charges 
\begin{equation}
\begin{tabular}{ll}
\hline
SU(5) mass operator & U$_{r}$(1) \\ 
\hline
\hline
$T^{\alpha \beta }F_{\alpha }$ & +2 \\ 
$F_{\alpha }\,\nu_{R}$ & -2 \\ 
$\nu_{R}^{t}\,\nu_{R}$ & -10 \\ 
$T^{\alpha \beta }\,T^{\gamma \delta }$ & -2\\
\hline
\end{tabular}
\end{equation}
We observe that the mass operators $T^{\alpha \beta }\,T^{\gamma \delta }$
and $F_{\alpha }\,\nu_{R}$ have the same charges, thus we expect that the
same SU(5) singlet is at the origin of their Yukawa interaction. As said, while
neutrinos have an approximate S$_{3}$ symmetry \cite{Caravaglios:2005gw}, 
the same symmetry is not observed in the up sector. 
If we embed SU(5) into E$_{6}$ we have an additional U$_{t}$(1),
$\mbox{E}_6\supset\mbox{SO}(10)\times\mbox{U}_t(1) \supset \mbox{SU}(5) \times \mbox{U}_r(1)\times \mbox{U}_t(1)$ 
where the 27 fundamental representation of E$_{6}$ contains an extra 
Standard Model singlet (27=1(4)+10(-2)+16(1)) and the table above becomes
\begin{equation}
\begin{tabular}{lll}
\hline
SU(5) mass operator & U$_{r}$(1) & U$_{t}$(1) \\ 
\hline
\hline
$T^{\alpha \beta }F_{\alpha }$ & +2 & +2 \\ 
$F_{\alpha }\,\nu_{R}$ & -2 & +2 \\ 
$\nu_{R}^{t}\,\nu_{R}$ & -10 & +2 \\ 
$T^{\alpha \beta }\,T^{\gamma \delta }$ & -2 & +2 \\ 
$F_{\alpha }\,x_{L}$ & +3 & +5 \\ 
$\nu_{R}^{t}\,x_{L}$ & -5 & +5 \\ 
$x_{L}^{t}\,x_{L}$ & 0 & +8\\
\hline
\end{tabular}
\end{equation}
The advantage here is that the 27 contains two standard model singlets 
that will play the role of right-handed neutrinos $\nu_{R}$ and $x_{L},$ 
and the Dirac mass operator $F_{\alpha }\,x_{L}$  has different
quantum numbers from all the others and in particular is different from $%
T^{\alpha \beta }\,T^{\gamma \delta }$ giving mass to the \ up sector. Thus
we explore the possibility that the fundamental lagrangian has a E$_{6}$
unifying gauge symmetry times a S$_{3}$ permutation symmetry of the three
fermion families that belong to the 27 of E$_{6}$. 
Now we have to choose the representation for the Higgs
 SU(2)$_{W}$ doublet which is a $S_3$ singlet. 
Now we have to decide to which E$_{6}$ representation we have
to assign the Higgs doublet. The 351$^\prime$ contains a SU(2)$_{W}$ doublet 
with (-3,-5) charges with respect the U$_{r}$(1)$\times $U$_{t}$(1) 
\cite{slansky}. Thus, if we  put the Higgs doublet in the 351$^{\prime }$, 
the Yukawa interaction for fermions at the tree level can be 
\begin{equation}\label{eq:2}
27_{i}^{\alpha }\,\ 27_{i}^{\beta }\,351_{\alpha \beta }^{\prime }
\end{equation}
where i=1,2,3 are family index and the 351$^{\prime }$ is symmetric under the exchange of $\alpha $ and $\beta $ the gauge symmetry indices. 
At the tree level of the fundamental high energy lagrangian, we
have just one Yukawa interaction 
$
g\,\ x_{iL}^{t}\,\ \nu_{iL}\,v
$
that comes from (\ref{eq:2})
since this is the unique U$_{r}$(1)$\times $U$_{t}$(1) gauge invariant 
operator. Thus there is only one Yukawa interaction in the fundamental
E$_6$ symmetric renormalizable Lagrangian that gives the Dirac neutrino
mass. The operator (\ref{eq:2}) does not introduce any mass neither for
quarks nor for charged leptons. We remember that this result is important
as explained above, since a Yukawa interaction $u_{R\,\ i}^{c}\,\
u_{L\,\ i}\,h_{0}$ (symmetric under S$_{3}$ family permutations) would give $%
m_{t}=m_{c}=m_{u}$ that is clearly unacceptable.
So, before the E$_{6}$ symmetry breaking, quark and charged
lepton yukawa couplings are zero, since they do not form a gauge invariant
operator with the Standard Model Higgs. The up quark yukawa operator 
is $T^{\alpha \beta }\,T^{\gamma \delta }H^{\sigma
}\,\varepsilon _{\alpha \beta \gamma \delta \sigma }$ and its charges are $%
(+5,+3)$. We need a SU(5) singlet with opposite U$_{r}$(1)$\times $U$_{t}$%
(1) to make an invariant operator. At first sight such a singlet is
contained both in the 78 and in the 650, it has the correct U(1) charges.
But we have shown \cite{Caravaglios:2005gf} that  in order to give a Yukawa 
coupling to the up quarks we have to write
an interaction 
$
27^{\alpha }\,27^{\beta }351_{\gamma \sigma }^{\prime }\Sigma _{\alpha \beta
}^{\gamma \sigma }
$
where $\Sigma _{\alpha \beta
}^{\gamma \sigma } $ is the irrep 2430.


\section*{Acknowledgments}
I would like to thank F. Caravaglios for the helpful discussion and the FP6
for the financial support.

\end{document}